\documentclass{PoS}
\usepackage{enumitem}

\title{Surface tension in the cold and dense chiral transition and astrophysical applications}

\ShortTitle{Surface tension in the cold and dense chiral transition and astrophysical applications}

\author{\speaker{Leticia F. Palhares}%
\\
 Institut de Physique Th\'eorique, CEA/DSM/Saclay, \\ Orme des Merisiers, 
       91191 Gif-sur-Yvette cedex, France \\
\\
       Instituto de F\'\i sica, Universidade Federal do Rio de Janeiro, \\
       Caixa Postal 68528, Rio de Janeiro, RJ 21941-972, Brazil\\
       E-mail: \email{leticia@if.ufrj.br}}

\author{Eduardo S. Fraga\\ 
         Instituto de F\'\i sica, Universidade Federal do Rio de Janeiro, \\
         Caixa Postal 68528, Rio de Janeiro, RJ 21941-972, Brazil\\
         E-mail: \email{fraga@if.ufrj.br}}

\abstract{
The surface tension of cold and dense QCD phase transitions has appeared recently as a key ingredient in different astrophysical scenarios, ranging from core-colapse supernovae explosions to compact star structure. If the surface tension is low enough, observable consequences are possible.
Its value is however not known from first-principle methods in QCD, calling for effective approaches. Working within  the framework of homogeneous nucleation by Langer, we discuss the steps that are needed to obtain the nucleation parameters from a given effective potential.
As a model for deriving the effective potential for the chiral transition, we adopt the linear sigma model with constituent quarks at very low temperatures, which provides an effective description for the thermodynamics of the strong interaction in cold and dense matter, and predict a surface tension of $\Sigma \sim 5$--$15~$MeV/fm$^{2}$, well below previous estimates. Including temperature effects and vacuum logarithmic corrections, we find a clear competition between these features in characterizing the dynamics of the chiral phase conversion.}

\FullConference{The many faces of QCD\\
                November 1-5, 2010\\
                Gent, Belgium}

\begin{document}

One of the many faces of QCD that has been attracting continuous attention in the last years is how the matter dictated by the fundamental theory of strong interactions behaves at extreme conditions of temperature and/or density. The fact that the energy scales for the QCD phase transitions are currently under reach of different heavy-ion colliders creates an intense demand of theoretical and phenomenological development.  Over the past decade RHIC has established the formation of a new state of matter, characterized by a clear fluid behavior at the partonic level, differing from the predicted picture of the na\"ive weakly-interacting quark gluon plasma \cite{Adams:2005dq}. LHC stretches the energy frontier also for heavy-ion collisions, bringing up the statistics and allowing for a careful study of fluctuations that might reveal the features of the initial state of the collisions. 

The richest structure in the high energy QCD phase diagram seems however to be deep within the region with an asymmetry between the particle and antiparticle content of the medium. In this direction many experimental efforts are dedicated to create matter with a finite baryon-chemical potential: the aim is the search for the phase boundaries in the QCD phase diagram and its properties. The accelerators however cannot reach the domain of very low temperature and high chemical potential, which is also the region where theorists have less guidance for what QCD predicts, since first-principle lattice calculations are not possible \cite{Laermann:2003cv}.

Astrophysical phenomena might however shed some light in this partially obscure region of the QCD phase diagram. Extremely high densities and relatively low temperatures are believed to exist in the core of ultracompact objects, so that it is probable that the QCD phase transitions play an important role in the structure and dynamics of such systems. Even though in general observations are much less controlable than laboratory experiments, the astronomical measurement techniques and instrumentation have much developed in the last decade and this might be the dawn of a precision era in which astrophysics really constrains nonperturbative QCD phenomena. One concrete example of astronomical observation that has reached the capability of constraining models for strong interactions in the nonperturbative regime is that of Ref. \cite{nature} in which the error bars of less than $5\%$ in the $2$ solar masses measurement rule out different equations of state.

The surface tension between partonic and hadronic phases of QCD seems to be a key quantity in connecting microscopic modeling of cold and dense QCD with astrophysical applications: recently it has been put forward that a low enough QCD surface tension should allow for possible observable consequences in different phenomena related to compact objects.
It was shown for instance that deconfinement can happen during the early post-bounce accretion stage of a core collapse supernova event, which could result not only in a delayed explosion but also in a neutrino burst that could provide a signal of the presence of quark matter in compact stars \cite{Sagert:2008ka}. However, as was discussed in detail in Ref. \cite{Mintz:2009ay} (see also Ref. \cite{Bombaci:2009jt} )  those possibilities depend on the actual dynamics of phase conversion, more specifically on the time scales that emerge. In a first-order phase transition, as is expected to be the case in QCD at very low temperatures, the process is guided by bubble nucleation (usually slow) or spinodal decomposition (``explosive'' due to the vanishing barrier), depending on how fast the system reaches the spinodal instability as compared to the nucleation rate \cite{reviews}. Nucleation in relatively high-density, cold strongly interacting matter, with chemical potential of the order of the temperature, can also play an important role in the scenario proposed in Ref. \cite{Boeckel:2009ej}, where a second (little) inflation at the time of the primordial quark-hadron transition could account for the dilution of an initially high ratio of baryon to photon numbers. Moreover, significantly different compact star structures are obtained if one considers the possibility of a layer of a quark-hadron mixed phase in the core of compact stars \cite{Kurkela:2010yk}. A key ingredient in all these scenarios is, of course, the surface tension, since it represents the price in energy one has to pay for the mere existence of an interface between quark and hadron phases. If the cost in energy is too high, quark-hadron mixed phases will not be favourable and similarly nucleation time scales will be too long for these astrophysical phenomena to take place.

Computing the surface tension for the QCD transitions with first-principle methods is a prohibitively complicated task.
It is a genuine nonperturbative quantity, in the sense that 
it cannot be obtained by a na\"ive expansion around one minimum of the effective action, and is also inaccessible by current first-principle lattice QCD simulations, due to the sign problem. One has therefore to resort to effective models. 
Estimates for the surface tension between a quark phase and hadron matter were 
considered previously in different contexts. In a study of the minimal interface between a 
color-flavor locked phase and nuclear matter in a first order transition, the authors of 
Ref. \cite{Alford:2001zr} use dimensional analysis and obtain $\Sigma\sim 300~$MeV/fm$^{2}$ 
assuming that the transition occurs within a fermi in thickness. Taking into account the effects 
from charge screening and structured mixed phases, the authors of Ref. \cite{Voskresensky:2002hu} 
provide estimates in the range of $50-150~$MeV/fm$^{2}$ but do not exclude smaller or larger 
values. These values of the surface tension are too high, rendering the nucleation of quark matter bubbles improbable in the context of compact objects.

%
%

We have estimated  \cite{Palhares:2010be} the surface tension for the cold and dense chiral phase transition within a well-established chiral model (namely, the linear sigma model with constituent quarks (LSMq)) and assuming the homogeneous nucleation scenario \cite{Langer:1967ax}. Our results indicate that the surface tension in the cold and dense regime can be considerably smaller than previous estimates, possibly allowing for the interesting observable astrophysical phenomena described above. 

This paper is organized as follows. In Section \ref{homnucl}, we delineate how, within the framework of homogeneous nucleation, it is possible to derive the nucleation parameters starting from an effective potential for the chiral order parameter $V_{\rm eff}(\sigma)$. Section \ref{EffM} contains a short presentation of the effective model we adopted for the QCD chiral phase transition and presents the results for the effective potential $V_{\rm eff}(\sigma)$ for three different cases. Our results for the surface tension are discussed in Section \ref{Res} and Section \ref{Conc} contains final remarks.

\section{How to estimate the surface tension?}\label{homnucl}

As stated above, with the aim of quantifying the surface tension associated with the cold and dense chiral phase transition in QCD, we will adopt the framework of homogeneous nucleation proposed by Langer \cite{Langer:1967ax}. It is based on the construction of a coarse-grained free energy functional $\mathcal{F}[\sigma]$ for the configuration of the order parameter field $\sigma$ from a given effective potential $V_{\rm eff}$ as follows:
\begin{eqnarray}
\mathcal{F}[\sigma]
&=&
4\pi\int r^2 dr\left[\frac{1}{2}\left(\frac{d\sigma}{dr}\right)^2+V_{\rm eff}(\sigma)\right]\,,
\label{Ffunc}
\end{eqnarray}
where spherical symmetry has been assumed.

\begin{figure}[h]
    \centering
        \includegraphics[width=0.35\textwidth]{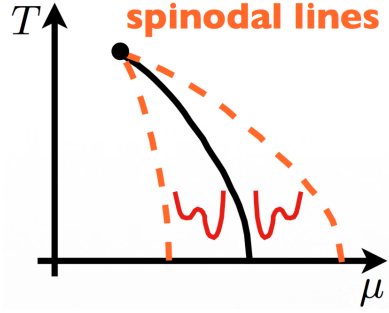}
\caption{Cartoon of the temperature {\it versus} chemical potential phase diagram with a first order critical line (solid) and the spinodal curves (dashed) defining the metastable region.}
\label{spinodals}
\end{figure}

\begin{figure}
\center
\begin{minipage}[h]{62mm}
\includegraphics[width=5.5cm]{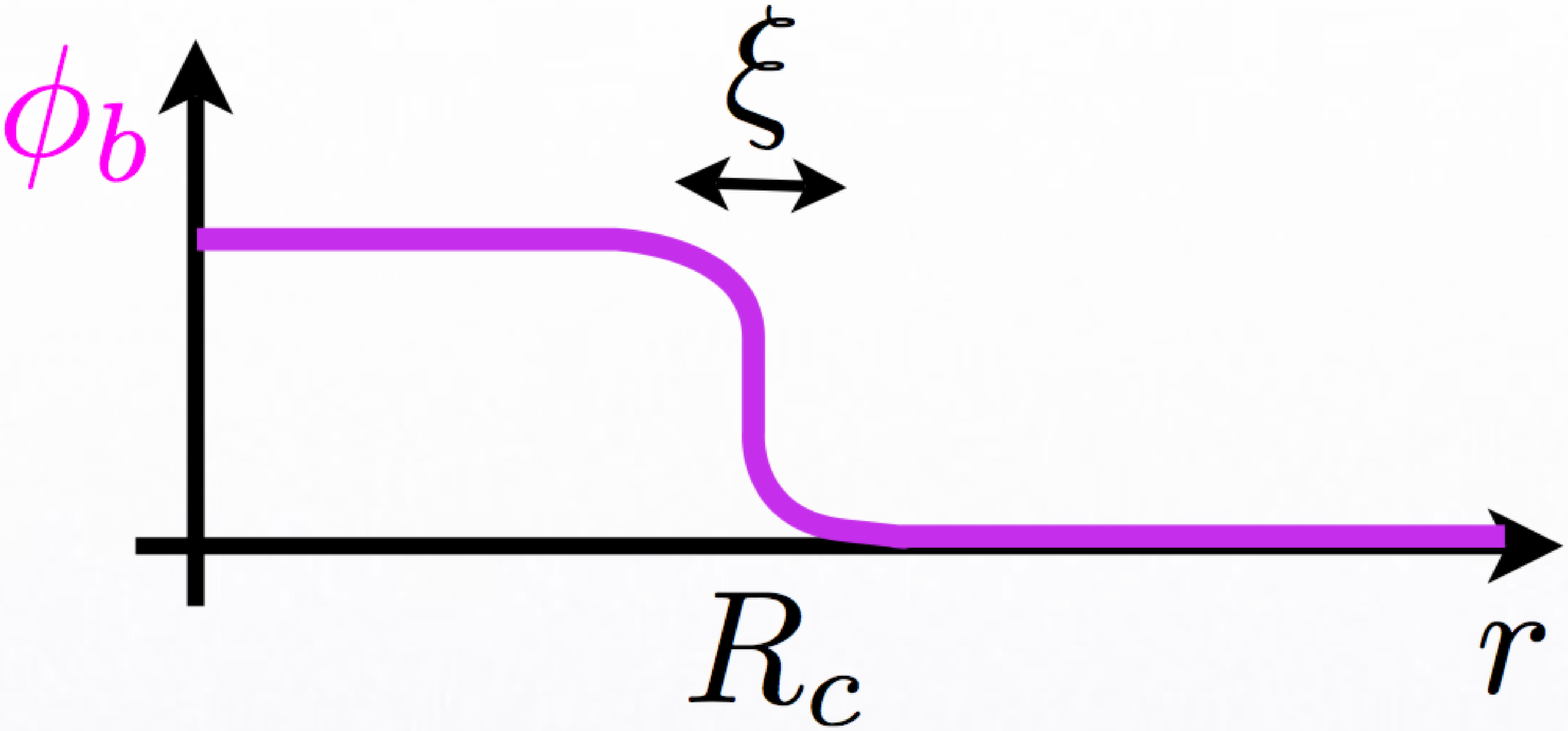}
\caption{Example of bubble profile.}
\label{bubble}
\end{minipage}
\hspace{.7cm} 
\begin{minipage}[h]{62mm}
\includegraphics[width=5.5cm]{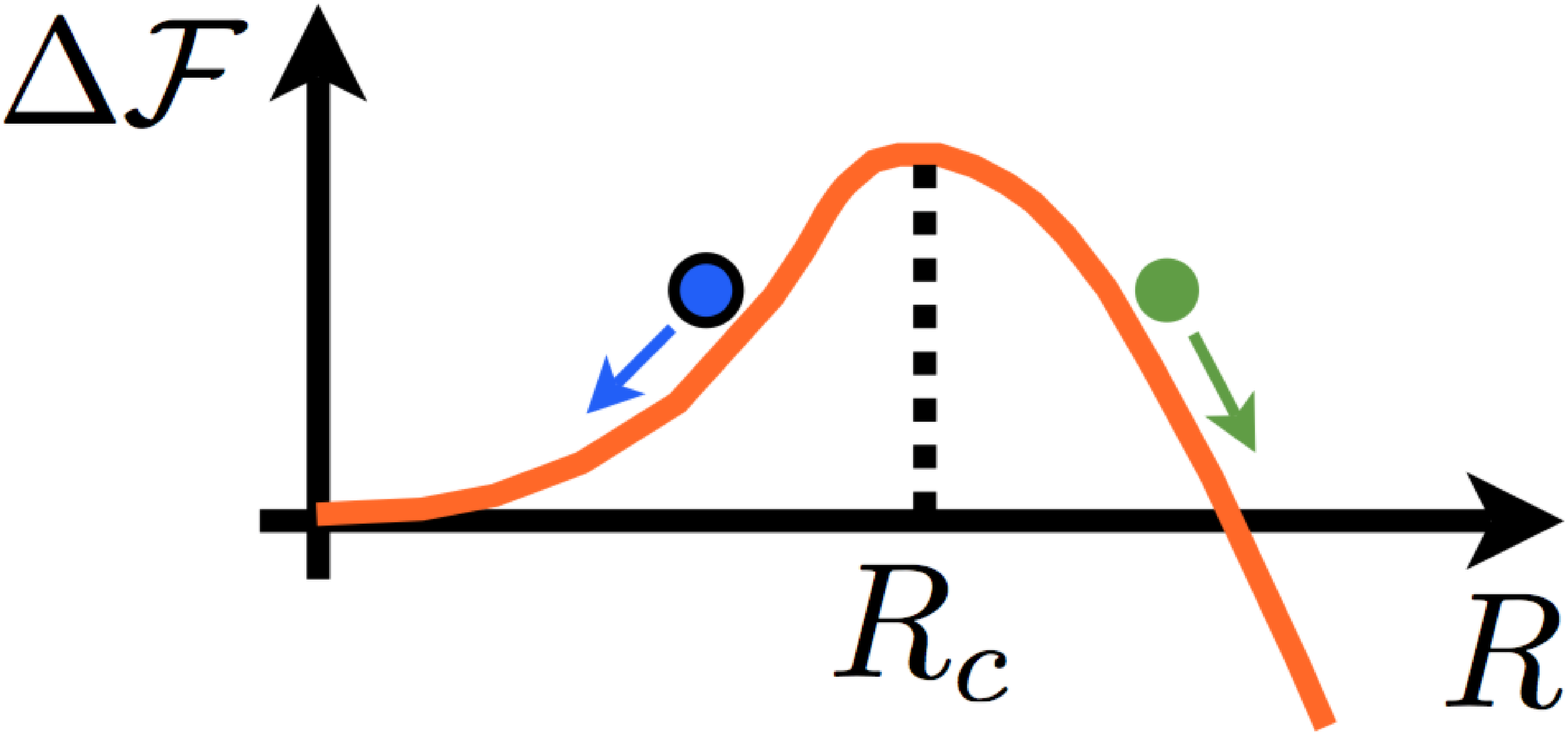}
\caption{Sketch of the difference of coarse-grained free energy between the configurations with and without a bubble of radius $R$.}
\label{DeltaF}
\end{minipage}
\end{figure}

Around the domain of temperature $T$ and chemical potential $\mu$ in which a first-order phase transition takes place, a metastable phase appears. More precisely, metastability manifests as the existence of (at least) two minima of the effective potential with a maximum in between inside the region defined by the two spinodal lines, as depicted in Fig. \ref{spinodals}. Within the so-called metastable region, the coarse-grained free energy functional in Eq. (\ref{Ffunc}) develops a nonuniform extremum configuration of the bubble type $\phi_b(r; R=R_c)$, connecting the metastable minimum to the absolute one (for an illustration cf. Fig. \ref{bubble}). This unstable nonuniform extremum is called the critical bubble and is intimately related to the dynamics of phase conversion via bubble nucleation. Differently from the uniform configuration sitting at the absolute minimum, the critical bubble is unstable, corresponding actually to a maximum of the free-energy $\mathcal{F}$.

If the system goes through an out-of-equilibrium stage in which the external parameters $T$ and $\mu$ are modified fast enough as compared to the time scales of the microscopic dynamics, then the system may get trapped in the metastable phase.
In this scenario, fluctuations (brought about by the interaction with a thermal or particle {\it reservoir})  may generate bubbles of the absolute minimum inside the metastable medium. These fluctuation bubbles will then evolve according to the competition between volume and surface terms, as dictated by the free-energy difference between the configurations with and without a bubble of radius $R$:
\begin{eqnarray}
\Delta\mathcal{F}
&\equiv&\mathcal{F}[\phi_b(r;R)]-\mathcal{F}[\phi_b(r;R=0)]
=
-\frac{4\pi}{3}R^3\Delta P+4\pi R^2\Sigma
\,,
\label{eqDeltaF}\end{eqnarray}
whose general form is sketched in Fig. \ref{DeltaF}. The volume term encodes the gain in energy resulting from the conversion to the true equilibrium state and is proportional to the pressure difference $\Delta P>0$ between the phases. On the other hand, the surface term accounts for the 
energy spent in the construction of a surface between the phases, being given essentially by the surface tension $\Sigma$. 
%

As illustrated in Fig. \ref{DeltaF}, the free-energy difference $\Delta \mathcal{F}$ shows a maximum associated with the critical bubble ($R=R_c$) and fluctuation-generated configurations containing bubbles with radii $R<R_c$ will tend to the uniform metastable vacuum in order to minimize the energy. For radii bigger than the critical radius, the minimization of energy implies that the droplet will grow and eventually complete the phase conversion. The critical radius goes to zero at the spinodal curves, where the metastable false vacuum becomes unstable (the barrier disappears) and the phase conversion occurs explosively via the spinodal decomposition process.

Finally, from Eq (\ref{eqDeltaF}) it is clear that the surface tension can be obtained via the knowledge of the effective potential $V_{\rm eff}$ and the critical bubble profile $\phi_b(r;R=R_c)$. Since the critical bubble itself is an unstable extremum of the coarse-grained action $\mathcal{F}$ in Eq. (\ref{Ffunc}), it is ultimately obtained from the effective potential $V_{\rm eff}$ as well. In the thin wall approximation (valid when the critical bubble has a thin interface region as compared to its radius, i.e. $\xi\ll R_c$; cf. Fig. \ref{bubble}) for a quartic effective potential, the solution $\phi_b(r;R=R_c)$ can be written analytically in terms of the Taylor coefficients of the potential.

\section{Effective model for the chiral phase transition}\label{EffM}

In order to estimate the chiral surface tension for cold and dense QCD matter, it suffices therefore, within the framework of homogeneous nucleation presented above, to provide an effective potential $V_{\rm eff}(\sigma)$ describing the chiral phase transition for low temperatures and high chemical potentials.

Here we consider the linear sigma model coupled with constituent quarks (LSMq) at zero or low temperature and finite quark chemical potential as a model for the thermodynamics of strong interactions in cold and dense matter. Since we are concerned with the chiral phase transition, we neglect the pions \cite{Scavenius:2000bb} and work with the following Lagrangian:
\begin{eqnarray}
{\cal L} &=&
 \overline{\psi}_f \left[i\gamma ^{\mu}\partial _{\mu} - m_q - g\sigma\right]\psi_f + \frac{1}{2}\partial _{\mu}\sigma \partial ^{\mu}\sigma
- U(\sigma)\,,
\end{eqnarray}
where
\begin{eqnarray}
U(\sigma)&=&\frac{\lambda}{4}(\sigma^{2} -
{\it v}^2)^2-h\sigma
\end{eqnarray}
is the self-interaction potential for the sigma meson, exhibiting both spontaneous and explicit breaking of chiral symmetry. The massive fermion fields $\psi_f$ ($f=1,\cdots,N_f=2$) represent the up and down constituent-quark fields. The scalar field $\sigma$ plays the role of an approximate order parameter for the chiral transition, being an exact order parameter for massless quarks and pions. 

We compute the effective potential                 
\begin{eqnarray}
V_{\rm eff}(\bar\sigma)
&=&
U(\bar\sigma)+\Omega^{\rm ren}(\bar\sigma) 
\end{eqnarray}
for the sigma condensate integrating over the quark fields and keeping quadratic fluctuations of the sigma field around the condensate. Our full thermodynamic potential     incorporates all corrections from the medium and vacuum fluctuations in the $\overline{\rm MS}$ scheme, including logarithmic and scale-dependent contributions from quark and sigma bubble-diagrams.
   Up to 1-loop order, the temperature- and chemical-potential-dependent contribution is that of an ideal gas ($N_c=3$ is the number of colors)
\begin{eqnarray}
\Omega_{{\rm med,Th}}^{(1)} &=&
T~\int\frac{d^3{\bf k}}{(2\pi)^3}
~\log\left[
1-{\rm e}^{-\omega_{\sigma}/T}
\right]
-\nonumber\\
&&
-2TN_f N_c\int\frac{d^3{\bf p}}{(2\pi)^3}
\left\{\log\left[
1+{\rm e}^{-(E_q-\mu)/T}
\right]+
\log\left[
1+{\rm e}^{-(E_q+\mu)/T}
\right]
\right\}
\,,
\end{eqnarray}
and the quantum vacuum term is given by:
\begin{eqnarray}
\Omega_{{\rm vac}}^{(1)} 
&=&
-\frac{M_{\sigma}^4}{64\pi^2}
\left[ \frac{3}{2}+\log\left( \frac{\Lambda^2}{M_{\sigma}^2} \right) \right] + N_f N_c~\frac{M_{q}^4}{64\pi^2}
\left[ \frac{3}{2}+\log\left( \frac{\Lambda^2}{M_{q}^2} \right) \right]
\,.
\end{eqnarray}

In the zero-temperature limit, the medium contribution reduces to:
\begin{eqnarray}
\Omega_{{\rm med}}^{(1)} =
- ~\frac{N_f N_c}{24\pi^2}
\left\{
2\mu p_f^3 -
3 M_{q}^2~\left[ \mu p_f-M_{q}^2\log\left( \frac{\mu+p_f}{M_{q}} \right) \right]
\right\}
\,,
\end{eqnarray}
with the effective masses:                                                                    
\begin{eqnarray}
M_q &\equiv& m_q + g \langle\sigma\rangle 
\,,\\
M_\sigma^2 &\equiv& 3\langle\sigma\rangle^2 -\lambda {\it v}^2
\,.
\end{eqnarray}

The parameters of the lagrangian are chosen such that the effective model reproduces correctly the phenomenology of QCD at low energies and in the vacuum, such as the spontaneous (and small explicit) breaking of chiral symmetry and experimentally measured meson masses.  The conditions for fixing the parameters are imposed on the vacuum effective potential, and therefore will be modified in the presence of vacuum logarithmic corrections. For more details on the parameter fixing procedure, the reader is referred to Ref. \cite{Palhares:2010be}.

In order to identify the role played by thermal effects and vacuum logarithmic corrections, we compare \cite{Palhares:2010be} three cases for the effective potential $V_{\rm eff}(\bar\sigma)
=U(\bar\sigma)+\Omega^{\rm ren}(\bar\sigma) $:
\begin{itemize}[leftmargin=*,label=]
\item {\bf (a)} including thermal corrections (with $T = 30~$MeV):
$\Omega^{\rm ren}
=\Omega_{{\rm med, Th}}^{(1)} $;

\item {\bf (b)} considering quantum vacuum terms: $\Omega^{\rm ren}
=\Omega_{{\rm vac}}^{(1)} +\Omega_{{\rm med}}^{(1)}$;

\item {\bf (c)} cold and dense LSMq: $\Omega^{\rm ren}
=\Omega_{{\rm med}}^{(1)}$.

\end{itemize}

\section{Results}\label{Res}

Let us now analyze the results for the surface tension in the LSMq. For more details, plots, discussion and results for other nucleation parameters the reader is referred to Ref. \cite{Palhares:2010be}.

For the zero-temperature chiral transition, the chiral surface tension $\Sigma$ assumes values between $4$ and $13\,$MeV$/{\rm fm}^2$, being well below previous estimates \cite{Alford:2001zr,Voskresensky:2002hu}.
We find that the nucleation parameters up to $T=10\,$MeV present variations within $\sim 10\%$ of the zero-temperature values. Here, motivated by applications to compact stars, we use $T=30\,$MeV, whenever thermal corrections are included.
The metastable region in the presence of quantum vacuum terms is $\sim 40\%$ bigger than in its absence, indicating possible large differences in the dynamics of the system, even though the critical chemical potential itself shifts only $\sim 2\%$.

\begin{figure}[h]
    \centering
        \includegraphics[width=0.5\textwidth]{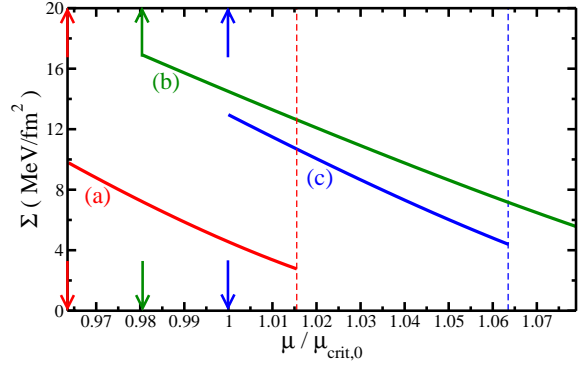}
\caption{Surface tension for quark-matter nucleation in the cold and dense LSMq (c), including vacuum terms (b) and thermal corrections (a).}
\label{Sigmavsmu}
\end{figure}

Fig. \ref{Sigmavsmu} shows a clear competition between vacuum and thermal corrections. As compared to the zero-temperature, classical computation (curve (c)), the establishment of an interface between quark matter droplets and the hadronic medium costs more with the incorporation of quantum corrections in the effective potential. On the other hand, finite temperature terms tend to push the surface tension down, reaching a minimum of only $\sim3\,$MeV$/{\rm fm}^2$, facilitating the surface formation.

\begin{figure}[h]
    \centering
        \includegraphics[width=0.5\textwidth]{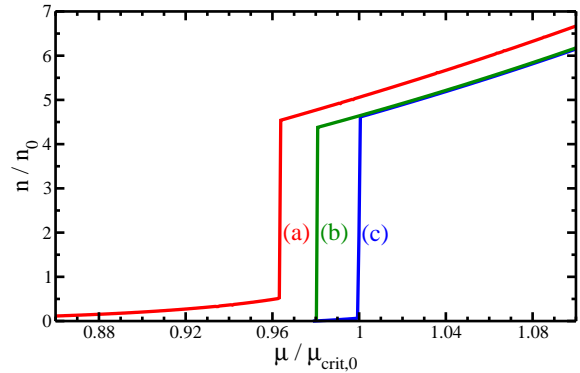}
\caption{Density in units of the nuclear saturation density ($n_0$) as a function of the chemical potential.}
\label{density}
\end{figure}

For astrophysical applications, it is sometimes useful to characterize the medium according to its density instead of chemical potential. The mapping between density and quark chemical potential within the cases we consider is plotted in Fig. \ref{density}, in which the discontinuities signal the first-order phase transition that restores chiral symmetry at high energies.

\vspace{0.5cm}

\section{Conclusions and perspectives}\label{Conc}

   We have quantified approximately the homogeneous nucleation in the linear sigma model with quarks in the $\overline{\rm MS}$ scheme at zero and low temperature and finite quark chemical potential, including vacuum and medium fluctuations. All the relevant quantities were computed \cite{Palhares:2010be} as functions of the chemical potential, the key function being the surface tension. 

   The inclusion of temperature effects and vacuum logarithmic corrections revealed a clear competition between these features in characterizing the dynamics of the chiral phase conversion, so that if the temperature is low enough the consistent inclusion of vacuum corrections could help preventing the nucleation of quark matter during the collapse process. In particular, we show that the model predicts a surface tension of $\sim5 -15\,$MeV$/{\rm fm}^2$, rendering nucleation of quark matter possible during the early post-bounce stage of core collapse supernovae. 

   The LSMq, however, does not contain essential ingredients to describe nuclear matter, e.g. it does not reproduce features such as the saturation density and the binding energy. Therefore, the results obtained here should be considered with caution when applied to compact stars or the early universe. It is an effective theory for a first-order chiral phase transition in cold and dense strongly interacting matter, and allows for a clean calculation of the physical quantities that are relevant for homogeneous nucleation in the process of phase conversion. In the spirit of an effective model description, our results should be viewed as estimates that indicate that the surface tension is reasonably low and falls with baryon density, as one increases the supercompression. First-principle calculations in QCD in this domain are probably out of reach in the near future. Therefore, estimates within other effective models would be very welcome.

\section*{Acknowledgements}
This work was partially supported by CAPES-COFECUB (project 663/10), 
CNPq, FAPERJ and FUJB/UFRJ.


\begin{thebibliography}{99}

\bibitem{Adams:2005dq} J.~Adams \textit{et al.} [STAR Collaboration], 
Nucl.\ Phys.\ A \textbf{757} (2005) 102. 

\bibitem{Laermann:2003cv}
E.~Laermann and O.~Philipsen,
Ann. Rev. Nucl. Part. Sci. {\bf 53} (2003) 163;
%
  S.~Hands,
  Prog.\ Theor.\ Phys.\ Suppl.\  {\bf 168} (2007) 253.

%
\bibitem{nature}
  P.~B.~Demorest, T.~Pennucci, S.~M.~Ransom, M.~S.~E.~Roberts, and J.~W.~T.~Hessels,
  Nature {\bf 467} (2010) 1081.


\bibitem{Sagert:2008ka}
  I.~Sagert {\it et al.},
  Phys.\ Rev.\ Lett.\  {\bf 102} (2009) 081101.
  
\bibitem{Mintz:2009ay}
  B.~W.~Mintz, E.~S.~Fraga, G.~Pagliara and J.~Schaffner-Bielich,
  Phys.\ Rev.\  D {\bf 81} (2010) 123012;
%
  J.\ Phys.\ G {\bf 37} (2010) 094066.
  
\bibitem{Bombaci:2009jt}
 I.~Bombaci, D.~Logoteta, P.~K.~Panda, C.~Providencia and I.~Vidana,
 Phys.\ Lett.\  B {\bf 680} (2009) 448.
  
\bibitem{reviews}
J. D. Gunton, M. San Miguel and P. S. Sahni, in 
\textit{Phase Transitions and Critical Phenomena} 
(Eds.: C. Domb and J. L. Lebowitz, Academic Press, London, 1983), v.~8.

\bibitem{Boeckel:2009ej}
  T.~Boeckel and J.~Schaffner-Bielich,
  Phys.\ Rev.\ Lett.\  {\bf 105} (2010) 041301
  [Erratum-ibid.\  {\bf 106} (2011) 069901].

\bibitem{Kurkela:2010yk}
  A.~Kurkela, P.~Romatschke, A.~Vuorinen and B.~Wu,
  arXiv:1006.4062 [astro-ph.HE].
  
\bibitem{Alford:2001zr}
  M.~G.~Alford, K.~Rajagopal, S.~Reddy and F.~Wilczek,
  Phys.\ Rev.\  D {\bf 64} (2001) 074017.

\bibitem{Voskresensky:2002hu}
  D.~N.~Voskresensky, M.~Yasuhira and T.~Tatsumi,
  Nucl.\ Phys.\  A {\bf 723} (2003) 291.

\bibitem{Palhares:2010be}
  L.~F.~Palhares and E.~S.~Fraga,
  Phys.\ Rev.\  D {\bf 82} (2010) 125018.

\bibitem{Langer:1967ax}
  J.~S.~Langer,
  Annals Phys.\  {\bf 41} (1967) 108
  [Annals Phys.\  {\bf 281} (2000) 941];
%
  Annals Phys.\  {\bf 54} (1969) 258.

\bibitem{Scavenius:2000bb}
  O.~Scavenius, A.~Dumitru, E.~S.~Fraga, J.~T.~Lenaghan and A.~D.~Jackson,
  Phys.\ Rev.\  D {\bf 63} (2001) 116003.




%
%
%
%
%
%
%
%
%
%
%
%
%
%
%
%
%
%
%
%
%
%
%
%
%
%
%
%
%
%
%
%
%
%
%
%
%
%
%
%
%
%
%
%
%
%
%
%
%
%
%
%
%
%
%




\end{thebibliography}
\end{document}